\documentclass[preprint]{aastex} 

\usepackage{amsmath,amssymb,amsfonts}
\usepackage{graphicx}
\usepackage{hyperref}
\usepackage{array}
\usepackage{cite}

\usepackage{verbatim}

\begin{document}

\title{Constraining the Rate of Primordial Black-Hole Explosions and Extra Dimension Scale using a Low-Frequency Radio Antenna Array}

\author{
   Sean E. Cutchin\altaffilmark{1},
John H. Simonetti\altaffilmark{1},
   Steven W. Ellingson\altaffilmark{2},
Amanda S. Larracuente\altaffilmark{3},
   \\
Michael J. Kavic\altaffilmark{3}
}
\altaffiltext{1}{Department of Physics, Virginia Tech, Blacksburg, VA 24061, USA}
\altaffiltext{2}{Bradley Department of Electrical and Computer Engineering, Virginia Tech, Blacksburg, VA 24061, USA}
\altaffiltext{3}{Department of Physics, Long Island University, Brooklyn, NY 11201, USA}
%
%
%

\setcounter{footnote}{0}
\renewcommand{\thefootnote}{\arabic{footnote}}

\begin{abstract}

An exploding primordial black-hole (PBH) may
produce a single pulse of electromagnetic radiation detectable at the low-frequency end of the radio spectrum.
Furthermore, a radio transient from an exploding PBH could be a signature of an extra spatial dimension.
We describe here an approach for searching for PBH explosions using a low-frequency radio antenna array, and as a practical example, the results of a such a search using the Eight-meter-wavelength Transient Array (ETA).
No compelling astrophysical signal was detected in $\approx 4$ hours of data implying an observational upper limit on the rate of exploding PBHs is $4.2 \times 10^{-7} \,\rm{pc}^{-3}\,\rm{yr}^{-1}$ for an exploding PBH with a fireball Lorentz factor of $10^{4.5}$ for the standard scenario of Page and Hawking. This rate limit is the strongest constraint yet set for PBH explosions with this fireball Lorentz factor. Observations ($\sim300$ hours) using the Arecibo Observatory were used to set a stronger constraint on the rate of PBH explosions for a fireball Lorentz factor of $10^{4.6}$ but the limit set by those observations for the fireball Lorentz factor considered here are less stringent by more than an order of magnitude.   
The limits considered here are applicable to exploding PBHs in the halo of the Galaxy. These observations also imply an upper limit of $2.0 \times 10^{-4} \,\rm{pc}^{-3}\,\rm{yr}^{-1}$ on the rate of PBH explosions in the context of certain extra dimension models as described by Kavic et al. This rate limit is for a fireball Lorentz factor of $10^{4.3}$ which corresponds to an extra dimension compactification scale of $3.0 \times 10^{-18}$~m.
\end{abstract}
\setcounter{footnote}{0}
\renewcommand{\thefootnote}{\arabic{footnote}}

\newpage

\section{Introduction}\label{sec:intro}

The study of black-holes underpins much of modern astrophysics. In 1975 Hawking suggested that black-holes emit energy like blackbodies with a temperature inversely proportional to the black-hole mass (\citet{Hawking:1975}). Thus, black-holes evaporate and the rate of evaporation is greater for smaller mass black-holes. While the evaporation rate for stellar mass black-holes is too low to imply practical observable consequences, small enough black-holes could completely evaporate on a time scale comparable to the age of the universe. Such small mass black-holes produced by the big bang --- primordial black-holes (PBHs) --- could now be reaching their endpoints and perhaps explode during their final moments. Observing these explosions would be of obvious cosmological and physical significance (\citet{Belotsky:2014kca, Khlopov:2008qy, Khlopov:1985jw}).

Page and Hawking predicted that an evaporating primordial black-hole would explode, releasing a final burst of energy upon reaching the QCD energy scale at $kT\sim$0.1~GeV (where $T$ is the temperature of the black-hole), producing a burst of gamma rays (\citet{Page:1976}). The Energetic Gamma-Ray Experiment Telescope (EGRET) set an upper
limit on PBH explosions of $<$ 0.059~pc$^{−3}$~yr$^{−1}$ (\citet{lehoucq2009new}), assuming the gamma-ray spectrum
peaks at about 250 MeV as discussed by Page and Hawking. 

However, there is another way to search for exploding PBHs which can yield better limits, as first pointed out by (\citet{Rees:1977}). A ``fireball'' of relativistic charged particles ejected by the explosion (e.g., electron-position pairs) would act as a superconducting, expanding shell, and expel the ambient interstellar magnetic field from a spherical volume centered on the exploding PBH, producing a radio pulse potentially detectable at large interstellar distances. Given the spectrum of radio emission derived by (\citet{Blandford:1977}) a low-frequency radio search can set better limits than gamma-ray searches.

A number of radio searches, mainly using single-dish telescopes, have yielded PBH explosion limits some of which are better than obtained by gamma-ray searches. In particular, one search by (\citet{Phinney:1979}) using $\sim300$ hours of observing time on the Arecibo telescope claimed a limit of $2 \times 10^{- 9}\,\rm pc^{-3} \,yr^{-1}$. However, this limit was based on the incorrect assumption of a flat radio pulse spectrum across the observing bandwidth. When the spectrum derived by (\citet{Blandford:1977}) is used the constraint becomes $1.1 \times 10^{- 8}\,\rm pc^{-3} \,yr^{-1}$ for a fireball Lorentz factor of $10^{4.6}$. The current work sets the strongest constraint on the PBH explosion rate limit for a Lorentz factor of $10^{4.5}$. The Arecibo observations noted above can be used to set a PBH explosion rate limit for a $5\sigma$ detection threshold of $4.8 \times 10^{- 6}\,\rm pc^{-3} \,yr^{-1}$ for this fireball Lorentz factor which is more than an order of magnitude less stringent that limit set here. As discussed below, the different values of the Lorentz factor correspond to different assumptions about either the particle spectrum at the QCD scale or the compactification scale of an extra spatial dimension, depending on the model being considered. Note that the rate limits noted above are for the traditional PBH explosion model due to Page and Hawking. It is also worth noting that these limits apply to a $\sim$arcminute beam of great depth, extending well beyond radius of the Galaxy. This is further discussed in section 5.

In this paper we demonstrate how a low-frequency radio antenna array instantaneously sensitive to a large fraction of the sky above the horizon can be used to conduct such a search that easily surpasses the limits set with single-dish telescopes. Unlike the Arecibo search, a search using a low-frequency antenna array such as described here would only be sensitive to exploding PBHs within our Galaxy or perhaps the galactic halo; it has been suggested that PBHs can be highly concentrated in galaxies, specifically in the galactic halo~(\citet{1996ApJ...459..487W}).

Low-frequency antenna arrays have been used to search for radio transients for many years (\citet{jelley1965radio, porter1965radio, balsano1996searching}). However, the commissioning of a new generation of low-frequency arrays (\citet{Ellingson:2007, 2012JAI.....150004T, stappers2011observing, tingay2013murchison}) in combination with the development of a new class of source models for radio transient production including novel PBH emission mechanisms (\citet{Kavic:2008, barrau2014fast}) suggest we reexamine our ability to use such arrays to search for transient signals. New developments in hardware and data processing techniques allow this new class of instruments to search for radio transients with greater sensitivity than ever before. Data from current arrays such as the first station of the Long Wavelength Array (LWA1) (\citet{2012JAI.....150004T}) and Low-Frequency Array for radio astronomy (LOFAR) (\citet{stappers2011observing}) can be searched for signals more effectively and with increased resolution than earlier instruments. We present the results of such a search we conducted using a small array of antennas which served as a pathfinder for the LWA1, the Eight-meter-wavelength Transient Array (ETA) (\citet{Ellingson:2007}). While the limit set by the observations presented is the strongest constraint available for a limited range of fireball Lorentz factors, further observations with the LWA1 or LOFAR would extend this work and set the most stringent limits to date for all possible fireball Lorentz factors.

In addition to the ``standard'' exploding PBH scenario of Page and Hawking, (\citet{Kavic:2008}) have recently pointed out that there is a second and observationally distinct explosion scenario, in this case for a PBH in the presences of an extra spatial dimension.  In short, an evaporating black-hole in the presence of an extra spatial dimension would undergo a black-string to 5-dimensional black-hole phase-transition that could produce an explosive transient event. This topological phase-transition explosion could also produce a detectable radio signal. We will compute possible limits obtainable using an antenna array for exploding PBHs in the standard scenario and the topological phase-transition scenario. The constraints on the scale of an extra dimension that are presented here are entirely novel. These constraints represent a truly modern method of using low-frequency arrays to search for exotic phenomena.

This paper is organized as follows:  Section~2 describes the two PBH explosion scenarios considered.  Section~3 discusses ETA, observations, and our data reduction procedure.  Section~4 presents observational limits on the PBH explosion rate.   Finally, we discuss the implications of our search in Section~5.

\section{Explosive PBH Scenarios}

\subsection{The Standard Scenario of Page and Hawking}
Hawking suggested that a black-hole emits energy like a blackbody~(\citet{Hawking:1975}) with the temperature defined as
\begin{equation}
\label{exactT}
T = \frac{\hbar c^3}{8\pi G k} \frac{1}{M},
\end{equation}
where $M$ is the mass of the black-hole. The emitted energy comes at the expense of the black-hole's mass,  and as its mass decreases, its temperature and mass loss rate increase. Since a black-hole radiates like a blackbody, the emitted power is
\begin{equation}
\label{power}
P = 4\pi R_s^2 \alpha(T) T^4,
\end{equation}
where $R_s$ is the Schwarzschild radius, and $\alpha(T)$ is proportional to the number of particle modes available at the temperature $T$.  At very low temperatures, where $kT$ is less than the energy of any particle of non-zero mass, the only particles that can be emitted are photons, and $\alpha$ is equal to the Stefan-Boltzmann constant, $\alpha = \sigma=\pi^2k^4/60\hbar^3c^2$. At higher temperatures a spectrum of photons and  particles of non-zero mass will be emitted.  The emitted energy comes at the expense of the black-hole's mass.
Thus
\begin{equation}
-\frac{dM}{dt}c^2=4\pi R_s^2 \alpha(T) T^4 \propto \frac{1}{M^2}.
\label{hi}
\end{equation}
Page and Hawking~(\citet{Page:1976}) described the explosion for a scenario in which all the remaining mass is emitted in a burst of energy $Mc^2$ near the QCD energy scale, $kT\sim0.1$~GeV.

\subsection{The Topological Phase-Transition Scenario}
(\citet{Kol:2002}) discussed a scenario whereby a compactified extra spatial dimension could produce a black-hole explosion. This scenario was considered in the context of PBH evaporation in (\citet{Kavic:2008}). Black-holes in four dimensions are uniquely defined by charge, mass, and angular momentum. However, with the addition of an extra spatial dimension, black-holes could exist in
different phases and undergo phase-transitions. For one toroidally compactified extra dimension, two possible phases are a ``black-string'' wrapping the compactified extra dimension, and a
5-dimensional black-hole smaller than the extra dimension.

When a PBH is larger in size than the extra dimension, it  wraps the extra dimension to form a black-string, and as the PBH evaporates it will eventually reach a radius comparable to the size of the extra spatial dimension.  The ``width" of the string is decreasing as the object evaporates while the ``length'' of the string is still wrapped around the extra dimension.  This situation is unstable (\citet{gregory1993black}) and the string should eventually snap, allowing the object to shrink to a size that can fit within the extra spatial dimension.  On ``snapping'' the black-hole releases a few percent of its mass-energy in an explosion of time scale $L/c$ where $L$ is the size of the extra spatial dimension.
The phase-transition from black-string to black-hole is expected to occur at a dimensionless (mass) parameter of $\mu \approx 0.07$, where $\mu=GM/Lc^2$. 
The mass-energy emitted by the object is
\begin{equation}
\eta Mc^2 = \eta\mu L c^4/G
\label{eq:energy}
\end{equation}
where $\eta\approx0.02$ is the efficiency~(\citet{Kol:2002}).

It is important to note that these two scenarios are mutually exclusive as the extra dimensional scenario would preclude a radio burst associated with the terminal outburst of the PBH as described in the standard scenario. This is because following the black-string to black-hole phase-transition evaporation takes place in 5-dimensional space and thus the power emitted is reduced as described by (\citet{Kol:2002}).  It should also be noted that a variety of models predict the existence of a single large extra spatial dimension. This includes universal extra dimension models (UED) which predict a compactification scale $\sim 10^{-18}$m as required by (\citet{Kavic:2008}) for radio pulse production.

\subsection{Radio Pulse Production}

\label{rpp}


Rees pointed out that searches for the coherent radio pulse that may be generated by the explosion of a black-hole would be more sensitive than searches for the emitted burst of Hawking radiation gamma rays (\citet{Rees:1977}). We discuss here a rough idea of what is happening in this radio pulse production.  The detailed properties of this pulse were discussed by (\citet{Blandford:1977}) and this analysis was later applied to the extra dimension scenario by (\citet{Kavic:2008}).  For the purpose of clarity and completeness we summarize their results below.

It is assumed that some substantial fraction of the mass-energy emitted in the explosion is in the form of electron-positron pairs. Nearly all of this energy, $mc^2$, is in the form of kinetic energy of the emitted particles (i.e., their Lorentz factors are $\gg1$). The particles form an outwardly expanding thin shell with an expansion speed $v$ (which is constant in this simple discussion) corresponding to a Lorentz factor $\gamma_f$.  The Lorentz factor of this expanding ``fireball'' is taken to be the typical Lorentz factor of the particles. In the case of electron-positron pairs, each pair will typically receive energy $kT$, thus each particle will end up with a Lorentz factor of
\begin{equation}
\label{Lorentz}
\gamma_f = \frac{\frac{1}{2}kT}{m_e c^2}
= \frac{\hbar c}{16\pi G m_e}\frac{1}{M}
\approx 10^5 \left( \frac{10^{11}~\text{g}}{M} \right)
\end{equation}
where $M$ is the mass of the black-hole at the moment of the explosion.

Since the expanding shell consists of charged particles, the ambient interstellar magnetic field energy is evacuated from the expanding bubble, ejected as electromagnetic radiation. In other words, the shell acts like an expanding, perfectly conducting sphere. The energy of the ambient magnetic field is boosted by $\gamma_f^2$ by being reflected off the expanding shell. The expansion ends when the expanding shell reaches a critical radius $R_c$ at which the ejected magnetic energy is equivalent to the energy emitted in the explosion.

Only a particular range of $\gamma_f$ can produce an electromagnetic pulse. The details are discussed in the papers by Rees and by Blandford. Simple arguments can be made for the upper and lower limits of $\gamma_f$ which illustrate a few of the processes taken into account in the papers. Below $\gamma_f \sim 10^5$ the energy emitted by the PBH goes primarily into sweeping up the ambient interstellar plasma, and not into an electromagnetic pulse; above $\gamma_f \sim 10^7$ the number of electron-positron pairs is insufficient to carry the fireball surface current necessary to expel the interstellar magnetic flux density.

The validity of the Rees model was questioned in~(\citet{1991ApJ...371..447M}). These authors use the standard model of particle physics as a basis for PBH emission at higher temperatures not the Hagedorn spectrum required by Rees and thus assume the black-hole emits quarks and gluons as predicted by standard QCD models. This analysis leads to the conclusion that a weak burst of $\gamma$-rays will result and not a coherent radio signal. A determination of which approach is correct depends chiefly on the physics of the QCD-scale.
It has been suggested that current experimental data is indicative of a Hagedorn spectrum at the QCD scale~(\citet{Broniowski:2004yh}) which would lend credibility to the scenario described by Rees while others disagree with this conclusion~(\citet{Cohen:2011cr}).  In the current work we simply constrain the density of PBHs assuming the Rees model as originally stated. Moreover, the extra dimension scenario also considered here has no dependence on QCD-scale physics and is not addressed in~(\citet{1991ApJ...371..447M}). Thus the objections raised do not apply.
The search described here yielded no positive detections and thus we are able to set constraints on both models: on the density of PBHs in the first case and on the density of PBHs and the compactification scale of extra spatial dimensions in the second.


Blandford worked out the details of the spectrum of an emitted radio pulse (\citet{Blandford:1977}).  The discussion below is based on his paper and is used here to analyze the topological phase-transition scenario. However, this analysis also applies to the standard Rees PBH explosion scenario with $\eta=1$. Using equation (\ref{Lorentz}) for the fireball Lorentz factor for a black-hole of mass $M$, the maximum energy that can be released is
\begin{equation}
Mc^2 = \frac{\hbar c^3}{16\pi Gm_e} \gamma_f^{-1}.
\end{equation}
For a topological phase-transition scenario, only a fraction of this mass-energy will be released.  Of the released energy, only a fraction will be put into charged particles. The energy released in charged particles (and ultimately into the electromagnetic pulse) is
\begin{equation}
\label{E}
E_{23} \approx \eta_{01} \ \gamma_{f5}^{-1}
\end{equation}
where $\eta_{01} = \eta/0.01$, $\gamma_{f5} = \gamma_f / 10^5$, and $E_{23}= E/10^{23}$J.  In equation (\ref{E}) the nominal value of 0.01 for the efficiency parameter $\eta$ reflects \textit{both} the expected few-percent efficiency of the mass-energy release by the phase transition, \textit{and} the assumption that $\sim50$\% of that energy is in the form of relativistic electron-positron pairs.\footnote{Blandford states that up to $\sim50$\% of the released energy might be in the form of electron-positron pairs but his final numerical results do not explicitly take account of this factor.  Benz and Paesold (\citet{Benz&Paesold:1998}) utilize Blandford's results without explicitly taking account of this factor.  Here, we have derived numerical results using $\eta\approx0.01$ which assumes that that 50\% of the released energy is in the form of relatvisitic electron-positron pairs, and ultimately goes into the electromagnetic pulse.}

Using equation (\ref{eq:energy}) we can relate the size of the extra dimension is to the Lorentz factor
\begin{equation}
L \approx \mu_{07}^{-1} \ \gamma_{f5}^{-1} \ 10^{-18} \ {\rm m},
\label{eq:gruga}
\end{equation}
where $\mu_{07} = \mu/0.07$.  This relation is used in section (\ref{sec:limit}) to constrain the size of a possible extra dimension in the context of the topological phase-transition scenario.

The fireball expansion time scale ($\approx R_c/c$), given Blandford's results, is
\begin{equation}
\Delta t_f = E_{23}^{1/3} \gamma_{f5}^{-2/3} b^{-2/3} \ 0.60 \text{\ s}
\end{equation}
which, combined with equation (\ref{E}) yields
\begin{equation}
\Delta t_f = \eta_{01}^{1/3} \gamma_{f5}^{-1} \ 0.60 \text{\ s}
\end{equation}
The critical frequency of the electromagnetic radiation produced by the fireball, following Blandford, is
\begin{equation}
\label{nuc}
\nu_c = E_{23}^{-1/3} \ \gamma_{f5}^{8/3} \ b^{2/3} \ 5.1\times10^9 \ {\rm Hz},
\end{equation}
where $b = \mathfrak{B}\sin\theta / 0.5\times10^{-9}$T is a magnetic field parameter, $\mathfrak{B}$ is the magnitude of the magnetic flux, and $\theta$ is the direction to the observer from the black-hole, in standard spherical coordinates, where the $z$-axis runs along the direction of the magnetic flux.  Combining equations (\ref{E}) and (\ref{nuc}) we get
\begin{equation}
\nu_c = \eta_{01}^{-1/3} \ \gamma_{f5}^3 \ b^{2/3} \ 5.1\times10^9 \ {\rm Hz}.
\end{equation}

From Blandford, the observed pulse energy spectrum (energy per unit frequency interval, per unit steradian) is
\begin{equation}
\label{I}
I_{\nu\Omega} = 1.4\times10^{12} \
E_{23}^{4/3} \
\gamma_{f5}^{-8/3} \
b^{-2/3} \
\left\vert F\left(\frac{\nu}{\nu_c}\right) \right\vert ^2
\ {\rm J\ Hz}^{-1}\ {\rm sr}^{-1},
\end{equation}
where
\begin{equation}
F(x) = \int_0^\infty dy\ y
\exp\left[ix\left(y+\frac{y^4}{2}+\frac{y^7}{7}\right)\right].
\end{equation}
The limiting forms of $\vert F(x)\vert^2$ are
\begin{equation}
\left\vert F(x) \right\vert^2 = \left\{ \begin{array}{ll}
0.615 \ x^{-4/7}-0.514 \ x^{-1/7}+0.027x^{2/7}+0.037 \ x^{5/7}+\mathcal{O}(x^{8/7}) & \textrm{if $x \lesssim 0.1$}\\
x^{-4}(1-75600 \ x^{-6}+\dots)   & \textrm{if $x \gtrsim 10$}.
\end{array} \right.
\label{spectrum2}
\end{equation}
Note that $I_{\nu\Omega}$ is the energy per unit frequency interval per steradian, and $I_{\nu\Omega}=2\pi I_{\omega\Omega}$ where $I_{\omega\Omega}$ is the energy emitted per \textit{angular} frequency interval, per steradian (the result Blandford produced).

Therefore, equations (\ref{E}) and (\ref{I}) yield
\begin{equation}
\label{spectrum}
I_{\nu\Omega} = 1.4\times10^{12} \
\eta_{01}^{4/3} \
\gamma_{f5}^{-4} \
b^{-2/3} \
\left\vert F\left(\frac{\nu}{\nu_c}\right) \right\vert ^2
\ {\rm J\ Hz}^{-1}\ {\rm sr}^{-1}.
\end{equation}

Observations of a pulse of specific Lorentz factor (i.e., observations of a specific observed spectrum) can be pinned to a particular efficiency factor if the distance to the explosion can be determined (e.g., through the dispersion measure, in the case of a radio pulse). Thus the ``standard'' explosion scenario ($\eta\approx1$) and the topological phase-transition scenario ($\eta\sim0.01$) can be distinguished.  This idea is discussed further by \citet{Kavic:2008}.  In the latter scenario, knowing $\gamma_f$ implies $L$.  A Lorentz factor of $10^5$ corresponds to an extra spatial dimension of $\approx10^{-18}$~cm, or an energy scale of $kT\approx(\gamma_f/10^5)$~0.1~TeV, the electroweak energy scale.  The mass of the black-hole at the moment of the phase-transition is about $10^{8}(\gamma_f/10^5)^{-1}$~kg.

\section{Observations and Data Reduction}

\subsection{General Considerations}

To calculate the observed signal-to-noise ratio, first assume the pulse is not dispersed or scattered.  For an explosion at distance $d$, detected by a single dipole of collecting area $A$, using a bandwidth $B$, and integration time $\tau$, the detected signal in units of energy, is
\begin{equation}
S =
\begin{cases}
I_{\nu\Omega} \frac{BA}{d^2} & \text{for } \tau = \Delta t \\
I_{\nu\Omega} \frac{BA}{d^2} \frac{\tau}{\Delta t} & \text{for } \tau < \Delta t.
\end{cases}
\end{equation}
This assumes the full bandwidth is detected coherently, and we have assumed the dipole matches the linear polarization of the arriving pulse (the pulse is $\approx$100\% linearly polarized since the interstellar magnetic field should be essentially uniform on the length scale $R_c$). We do not consider the case $\tau>\Delta t$ since the general data analysis procedure is to start with the highest temporal resolution time series and then smooth it, increasing $\tau$ until it matches $\Delta t$, yielding the highest signal-to-noise ratio for a pulse for this matched situation.

The rms temperature measurement for a radiometer is given by the so-called ``radiometer equation''
\begin{equation}
\sigma_T \approx \frac{T_{sys}}{\sqrt{B\tau}}
\end{equation}
for system temperature $T_{sys}$, bandwidth $B$, and integration time $\tau$.  The power in noise is $k\sigma_{T}B$.  If we measure the pulse signal in units of energy, the corresponding noise should also be in energy units.  As discussed by (\citet{Meikle&Colgate:1978}) the noise $N$ in units of energy, for integration $\tau$, is the noise power multiplied by $\tau$, or
\begin{equation}
N\approx\frac{kT_{sys}B}{\sqrt{B\tau}} \tau \approx kT_{sys}\sqrt{B\tau}.
\end{equation}
Here the full bandwidth is used in the measurement of the signal. $T_{sys}$ depends on the LST of the ETA observation, and varies between about 6000~K and 10,000~K.

The signal-to-noise ratio is therefore
\begin{equation}
\frac{S}{N} \approx
\begin{cases}
I_{\nu\Omega} \frac{BA}{d^2} \frac{1}{kT_{sys}\sqrt{B\tau}} & \text{for } \tau \ge \Delta t \\
I_{\nu\Omega} \frac{BA}{d^2} \frac{1}{kT_{sys}\sqrt{B\tau}} \frac{\tau}{\Delta t} & \text{for } \tau < \Delta t.
\end{cases}
\end{equation}
The best $S/N$ is obtained for $\tau=\Delta t$, which is when $\tau$ is as small as possible without cutting off some signal from the integration. For $N_{dipoles}$ independent dipoles, if the resulting time series are added together incoherently, the final signal-to-noise is $\sqrt{N_{dipoles}/2}$ better than the single dipole result (the $\sqrt{1/2}$ comes in since the source is linearly polarized).  In practice our $\tau$ will be set during the data analysis, as we integrate the data over a series of time samples.  We will be trying a range of $\tau$ in an attempt to match $\tau$ to the duration of a pulse buried in the data.

\subsection{Interstellar Dispersion, Scattering and Signal-to-Noise Ratio}

Both interstellar dispersion and interstellar scattering may be important in determining the observed pulse duration.  In general, the observed pulse duration is
\begin{equation}
\Delta t_{obs} \approx \sqrt{\Delta t_D^2 + \Delta t_{scatt}^2 }
\label{dur}
\end{equation}
where $\Delta t_{obs}$ is the observed pulse width, $\Delta t_D$ is the pulse broadening due to dispersion, and $\Delta t_{scatt}$ is the pulse broadening due to scattering. Strictly, equation (\ref{dur}) should include the intrinsic pulse width and observed pulse width contribution from an error in the observed dispersion measure ($DM$), both added in quadrature within the square root. However, the intrinsic pulse width is negligibly small, being on the order of a nanosecond (it is the inverse of the critical frequency given by equation (\ref{nuc})), and the $DM$ error can be reduced by decreasing the $DM$ step size in the $DM$ search. So, these two terms are neglected in this discussion.

We will first consider the effect of interstellar dispersion. Break the bandwidth $B$ up into $n$ frequency channels, each of width $\Delta \nu = B/n$.  We will incoherently de-disperse a pulse by shifting and adding these channels together, compensating for the relative delay between each channel.
For a dispersed pulse, after appropriate de-dispersion and summing of channels, we obtain the same $S/N$ as in the no dispersion case, as long as the pulse is not smeared in any one channel to a time duration longer than a time sample.  In other words, for best signal-to-noise, match $\tau$ to the dispersed pulse duration.

At some particular frequency channel, of width $\Delta\nu$, a pulse of originally infinitesimal duration will be spread in time to duration
\begin{align}
\Delta t_D &= \left( \frac{dt}{d\nu} \right)_D  \Delta\nu \\
&= 0.15 \text{s\ } DM \left( \frac{\nu}{\text{38\ MHz}} \right)^{-3}
\left( \frac{\Delta\nu}{\text{MHz}} \right) \\
&= 0.11 \text{s\ } \left( \frac{DM}{\text{50\ pc\ cm$^{-3}$}} \right)
\left( \frac{\nu}{\text{38\ MHz}} \right)^{-3}
\left( \frac{\Delta\nu}{\text{15\ kHz}} \right)
\end{align}
where $DM = \int n_e dl$ is the line of sight integral of the free electron density $n_e$.
After de-dispersion, and summing,
\begin{equation}
\frac{S}{N} \approx
I_{\nu\Omega} \frac{BA}{d^2} \frac{1}{kT_{sys}\sqrt{B\Delta t_{D}}}
\end{equation}
assuming the integration time is matched to $\Delta t_{D}$, yielding the best signal-to-noise ratio.

Interstellar scattering can also broaden the pulse.  
From (\citet{Cordes&McLaughlin:2003}) we can obtain a model of pulse scatter broadening as a function of $DM$. Matching $\tau$ to the final $\Delta t$ provides the largest $S/N$, of course.  But $\tau$ may have to be large. The scatter broadened pulse duration is approximately
\begin{equation}
\log\left( \frac{\Delta t_{scatt}}{\text{seconds}} \right) \approx
-9.72 + 0.411 \log\ DM + 0.937(\log DM)^2 - 4.4\log \nu_{\text{GHz}} \pm 0.65
\end{equation}
where $DM$ is in units of pc~cm$^{-3}$. For a $DM\sim30$~pc~cm$^{-3}$ the scatter broadening time is on the order of a few tenths of second at 38 MHz.
At 38~MHz the diffractive scintillation (twinkling) is ``quenched'' (smoothed out) in our $\Delta\nu$ frequency channels, because the so-called ``diffractive scintillation bandwidth'' is much less than $\Delta\nu=B/n$, so only scatter broadening of a pulse will affect $S/N$. The diffraction scintillation bandwidth is $\Delta\nu_{diff}\sim 1/2\pi\Delta t_{scatt}$ \citep{1998ApJ...507..846C}, or $\sim1$~Hz for a $\Delta t_{scatt}\sim$ a few tenths of a second; therefore $\Delta\nu_{diff}$ is much less than $\Delta\nu$ (7.32~kHz, for our ETA observations, see below).


Finally, matching the integration time $\tau$ to the pulse duration, and adding the signals incoherently from $N_{dipoles}$ independent dipoles, the resulting signal-to-noise ratio is
\begin{equation}
\frac{S}{N} \approx
I_{\nu\Omega} \frac{BA}{d^2} \frac{1}{kT_{sys}\sqrt{B\Delta t_{obs}}} \sqrt{\frac{N_{dipoles}}{2}}.
\label{eq:SN}
\end{equation}
For an array of dipoles, one could add the dipole signals coherently, i.e., to form a beam, but that beam would necessarily cover a smaller part of the sky than is observed by each dipole. Given that we do not know, in advance, the direction of the incoming radio pulse, the best search scenario should use the widest beam possible. Thus we chose to add the signals incoherently, producing equation (\ref{eq:SN}).

\subsection{ETA Observations}

Data for the current work were collected with the ETA in 5 observing sessions between 18~November and 03~December, 2007.
The observing sessions lasted between 30~minutes and 1~hour.
For a full description of ETA\footnote{ETA is a joint project between the Bradley Department of Electrical and Computer Engineering and the Department of Physics at Virginia Tech (\citet{Ellingson:2007}).} see (\citet{Ellingson:2007}) and (\citet{KshitijaETA:2009ps}).
ETA was designed to provide roughly uniform sensitivity over most of the visible sky.
It consisted of 12 dual-polarized dipole antennas that operated in the band 29 -- 47 MHz.
All data sets were taken with ETA at 38~MHz frequency with a 3.75~MHz processed bandwidth.
ETA parameters relevant to this observation on listed in Table \ref{tb:obsparam}. Note, in particular, that only 4 dipoles were used in these observations.
Following the analysis presented above we find the flux density that will produce $S/N\ge5$ is
\begin{equation}
f_\nu \gtrsim 5 \frac{kT_{sys}}{A \sqrt{B \Delta t_{pulse}}} \frac{1}{\sqrt{N_{dipoles}}}.
\end{equation}
For the observing parameters in Table \ref{tb:obsparam}, the required pulse flux density is given by
\begin{equation}
f_\nu \gtrsim 570 \text{\ Jy} \left( \frac{\Delta t_{pulse}}{1 \text{\ s}} \right)^{-1/2}.
\end{equation}

ETA is strongly noise limited by
galactic synchrotron emission, as opposed to any internal or terrestrial noise, as demonstrated by (\citet{Ellingson:2007, KshitijaETA:2009ps}).
In addition, (\citet{Ellingson:2007}) demonstrated the basic functionality of ETA obtaining observations that showed the expected frequency dependence of the total received power, and the diurnal variation of total received power from the Galaxy, extrapolating from the 408 MHz sky survey of (\citet{Haslam:1982}).

Removal of the instrumental frequency response yields an estimate of the power spectral density $S_{sky}$ at the antenna terminals, which can be converted into a temperature via $S_{sky} = e_{r} kT_{sky}\Delta \nu$, where $k$ is Boltzmann's constant, $e_{r}$ is the loss due to the finite
conductivity of the materials used to make the antenna as well as the
absorption by the imperfect (nonperfectly-conducting) ground and $\Delta \nu$ is the spectral resolution bandwidth~(\citet{Ellingson:2007}).
The expected value of $T_{sky}$ is given by the Rayleigh-Jeans approximation,
\begin{equation}
T_{sky} = \frac{1}{2k} \, I_{\nu} \, \frac{c^2}{\nu^2}.
\label{eq:Tsky}
\end{equation}
From observations of the Galactic polar region carried out by (\citet{1979MNRAS.189..465C}) we assume
\begin{equation}
I_{\nu} = I_{g} \left(\frac{\nu}{\rm{MHz}}\right)^{-0.52} + I_{eg}\left(\frac{\nu}{\rm{MHz}}\right)^{-0.80},
\label{eq:Inu}
\end{equation}
where $I_g = 2.48 \times 10^{-20}$ and $I_{eg} = 1.06 \times 10^{-20}$.

\subsection{Data Reduction}\label{sec:analysis}
We performed a single-dispersed-pulse search as described by (\citet{cutchin2011search}) and (\citet{Cordes&McLaughlin:2003}).
As discussed in detaill above, a radio pulse is dispersed as it propagates through the ionized interstellar medium (ISM) or the intergalactic medium (IGM).  The magnitude of this effect is quantified by the dispersion measure (DM), which is the integral of the electron column density along the path of propagation.  For example, a signal from a pulsar out of the plane of the Galaxy with a DM $\sim30\,\rm pc\,cm^{-3}$ would experience a delay of $\sim$90~seconds at 38~MHz.  The delay scales as the inverse square of the observing frequency, making it greater at lower frequencies.  The DM of a new source or transient is not known a priori, therefore a dedispersion search is performed over many candidate DMs.
The data were incoherently dedispersed across 1153 trial DMs in the range 10 -- 100~$\rm{pc}\,\rm{cm}^{-3}$ with a DM spacing $\delta DM = 0.002$~DM, and searched over multiple assumed pulse widths in the range 10~ms~--~2~s.
Dedispersion searching has proven effective for the detection of giant pulses, rotating radio transients (RRATs)~(\citet{McLaughlin:2006}), and the recently detected fast radio bursts (FRBs)~(\citet{2013Sci...341...53T}).
The dedispersed time series are smoothed with a range of integration times to search for pulses matched in duration to the integration time, which would yield the greatest signal-to-noise ratio for a pulse.

\begin{figure}[]
\centering
\includegraphics[width=110mm]{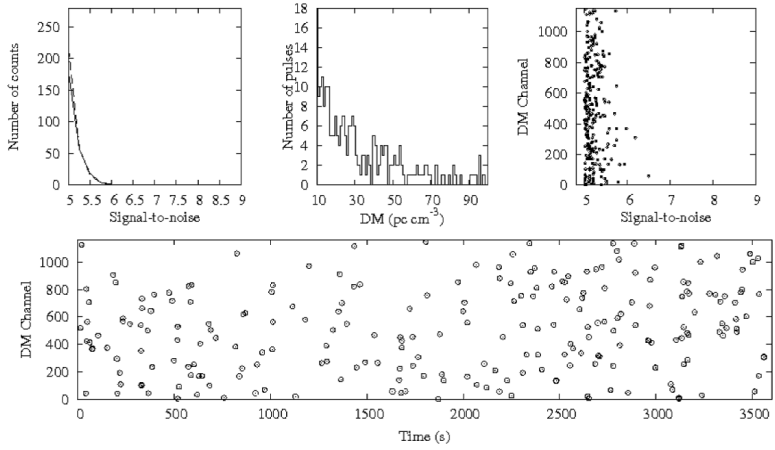}
\caption{Shown here is a representative plot of one hour of the observations conducted with the ETA. The upper left panel is histogram of S/N values.  The dashed curve represents the number of pulses above a S/N = 5 due to noise alone, which is expected to follow a Gaussian distribution in the absence of any signal(s). The number of recorded pulses (solid line) does not clearly deviate from the expected number for a Gaussian process. The upper center plot is a histogram of the number of pulses across the DM values that were searched. The upper right plot shows no correlation between high S/N pulses and a particular DM channel. The bottom plot is a time-series across the DM ranges searched. The size of the circular data points indicates the signal strength for each pulse. }  
\label{fig:Cordes}
\end{figure}

During these observations, no astrophysical transients were detected above the 5$\sigma$ level.
The frequency of spikes in the time series at the 5$\sigma$ level was consistent with that expected from Gaussian noise. A representative plot of one hour these observations confirming the expected Gaussian dependence of the background and the lack of astrophysical signals is shown in Figure (\ref{fig:Cordes}).

\section{Observational Limits on PBH Explosion Rate}\label{sec:limit}

\begin{figure}
\centering
\includegraphics[width=90mm]{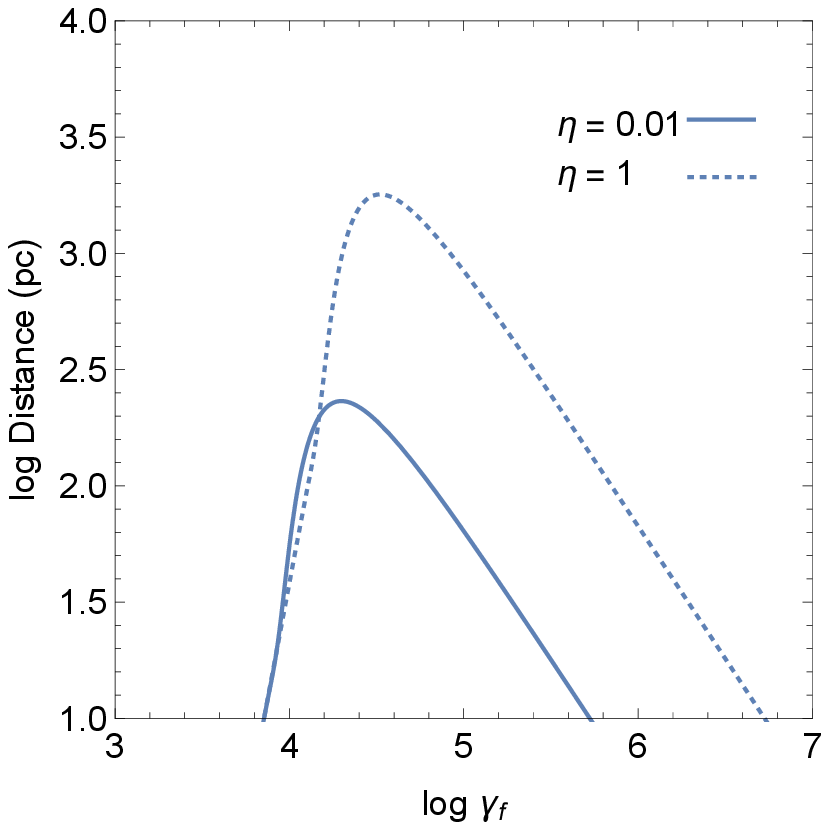}
\caption[$\log \gamma_f$ vs. $d_{max}$]{The distance to which a PBH explosion of $\eta = 0.01$ (solid curve) and $\eta = 1$ (dashed curve) for the range of possible fireball Lorentz factors, is shown for the ETA with 4 dipoles. The detection threshold used is $S/N >5$. The maximum detection distance is obtained for $\gamma_f=4.5$.}
\label{fig:gamma_dmax}
\end{figure}

The signal-to-noise ratio can now be calculated for various fireball Lorentz factors ($\gamma_f$) using the observing parameters in Table~\ref{tb:obsparam} and equation (\ref{eq:SN}).  
To give an example result for discussion we use source parameters of $\eta=0.01$ (or $\eta_{01}=1$), an interstellar magnetic flux parameter of $b=1$, and $DM=30$~pc~cm$^{-3}$, which is appropriate for lines of sight with high galactic latitudes (which would have the most beneficial $S/N$).
Then, given the values in Table \ref{tb:obsparam}, the largest signal-to-noise ratio is for an explosion with
\begin{equation}
\gamma_f = 10^{4.3}
\end{equation}
or $\gamma_{f5} = 0.199$. (See Figure (\ref{fig:gamma_dmax}).)
In that case the emitted pulse energy is $5.01 \times 10^{23}$~J, the critical frequency is $40$~MHz, and the intrinsic pulse duration before any dispersion or scattering is about $25$~ns.
%
The resulting emitted pulse energy per frequency interval, per steradian at 38 MHz is
\begin{equation}
I_{\nu\Omega} = 1.67 \times 10^{14} \,\rm{J}\,\rm{Hz}^{-1}\,\rm{sr}^{-1}.
\end{equation}
The observed pulse duration considering only dispersion broadening of the final detected pulse, using the nominal parameters for the PBH explosion, is $\Delta t_{D} = 0.066$~s. Scatter broadening alone would produce an observed pulse duration of $\Delta t_{scatt} \approx 0.151$~s.  Therefore the observed pulse duration including the combined interstellar broadening effects is $\Delta t_{obs} = \sqrt{\Delta t_{D}^2 + \Delta t_{scatt}^2} \approx  0.165 \,\rm{s}$.
Using equation~(\ref{eq:SN}) and the observation and source parameters noted above
we find that
a PBH explosion can be detected with $S/N = 5$ to a distance of $\sim 230 \,\rm{pc}$ and $1.8 \,\rm{kpc}$ for  $\eta = 0.01$ (the topological phase-transition) and  $\eta = 1$ (the standard scenario), respectively.

The solid angle searched by an ETA dipole at any moment is $\Omega \approx 2.6 \,\rm{sr}$ at 38 MHz~(\citet{KshitijaETA:2009ps}).  Therefore, the volume searched at any moment is $V = (1/3) d_{max}^3 \Omega$~(\citet{Phinney:1979}), where $d_{max}$ is the maximum distance to which such an explosion can be detected.  If the observations are conducted over a time period $T$, then the upper limit to the rate $r$ of PBH explosions, given no detections, is~(\citet{Phinney:1979})
\begin{equation}
  r \approx \frac{1}{VT} = \frac{1.15}{d_{max}^3 T}.
\label{rate}
\end{equation}
For 4.15 hours of ETA observing
an upper limit on PBH explosions can be set at
\begin{equation}
r \approx  \left\{
 \begin{array}{rl}
    2.0 \times 10^{-4} \, \rm{pc}^{-3} \,\rm{yr}^{-1} & \rm{for} \,\eta = 0.01, \gamma_f = 10^{4.3} \\
    4.2 \times 10^{-7} \, \rm{pc}^{-3} \,\rm{yr}^{-1} & \rm{for} \, \eta = 1, \gamma_f = 10^{4.5}.
  \end{array} \right.
 \label{eq:ratelimit}
\end{equation}

\begin{figure}
\centering
\includegraphics[width=90mm]{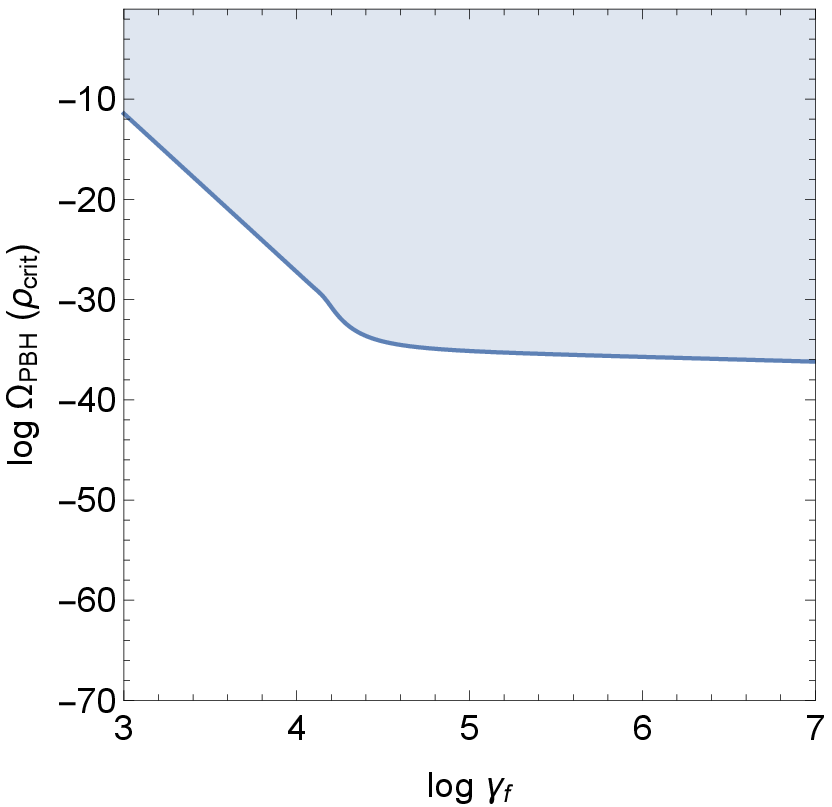}
\caption{The shaded region shows the section of parameter space defined by $\Omega_{PBH}$ and $\gamma_f$  which is excluded by our search.}  
\label{fig:GL}
\end{figure}
\begin{figure}
\centering
\includegraphics[width=90mm]{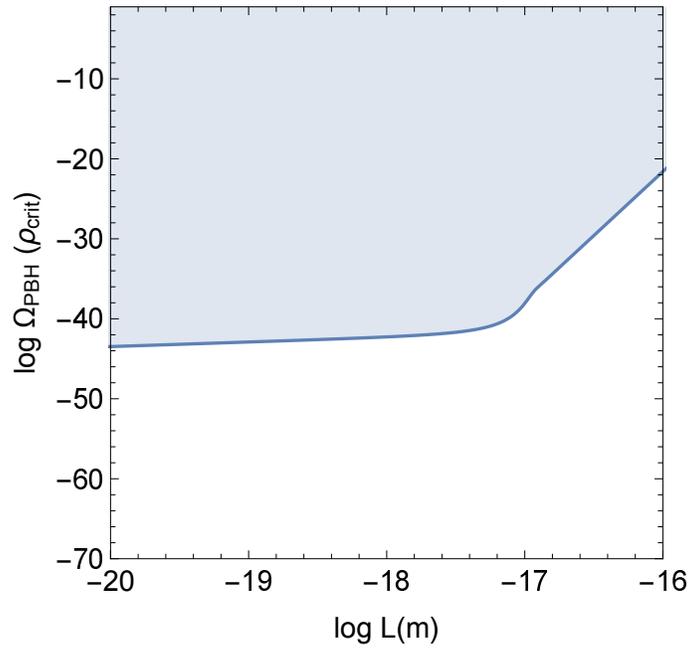}
\caption{The shaded region shows the section of parameter space defined by $\Omega_{PBH}$ and $L$ which is excluded by our search.  This constrains the allowed compactification scale in extra dimensional models.  The excluded region would apply to extra dimensional models that could be demonstrated to also lead generically to PBH production at the requisite mass scale when applied to early cosmology.}
\label{fig:OL}
\end{figure}

These rate limits can be converted into limits on the current density of PBHs, $\Omega_{PBH}$, as a fraction of the critical density, $\rho_{crit}$.
This can be done by relating the current PBH mass spectrum, $dN/dVdM$, to the current rate of PBH explosions, $dN/dVdt$, using
\begin{equation}
\frac{dN}{dVdM} = \Big( \frac{dN}{dVdt}\Big) \Big( \frac{dt}{dM}\Big),
\end{equation}
where $dt/dM$ can be found by inverting equation~(\ref{hi}).
(\citet{2005astro.ph.11743C}) has shown that the mass spectrum is directly related to $\Omega_{PBH}$ as
 \begin{equation}
 \frac{dN}{dVdM} = (\alpha-2) (M/M_\ast)^{-\alpha}M_\ast^{-2}\Omega_{PBH}\rho_{crit},
 \end{equation}
where $M_\ast$ is the current lower cut-off in the mass spectrum due to PBH evaporation and $\alpha=5/2$ for a radiation equation of state during the formation stage of PBHs.
As described above the Lorentz factor of the particles emitted by the PBH directly determines the volume searched and is thus directly related to the rate limit which can be set by our observations.
Thus the relevant range of values for $\gamma_f$ are related to different limits on $\Omega_{PBH}$.
Figure (\ref{fig:GL}) shows the limit on $\Omega_{PBH}$ plotted verses $\gamma_f$ showing the region excluded by our observations.
In the same way, our PBH rate density limit in the extra dimension scenario can be converted to a limit on $\Omega_{PBH}$.
Moreover, as noted in equation~(\ref{eq:gruga}), the Lorentz factor is directly related to the compactification scale~$L$.
Thus Figure (\ref{fig:OL}) shows our limit on $\Omega_{PBH}$ plotted verses $L$ showing the region excluded by our observations.

 \begin{table}
 \caption{Parameters of the ETA search.}
\centering
 \begin{tabular}{ccc}
 Parameter & Value \\ \hline \hline
 Observing frequency ($\nu$) & 38 MHz  \\
 Single dipole collecting area ($A$) & 18.8 $\rm{m}^2$\\
 Beam solid angle ($\Omega$) & $2.6 \,\rm{sr}$\\
 Number of Dipoles ($N_{dipoles}$)&4\\
 Sytem Temperature ($T_{sys}$) & 6000 K\\
 Processed Bandwidth (B) & 3.75 MHz\\
 Minimum integration time ($\tau_{min}$) & 9.97 ms\\
 Frequency channel width ($\Delta \nu$) & 7.32 kHz\\ \hline
 \end{tabular}
 \label{tb:obsparam}
 \end{table}
\section{Discussion}
The observations conducted by ETA, while lacking a positive detection of a radio transient, enable setting constraints on the rate of PBH explosions and the density of PBHs.  These constraints in turn inform our understanding of the early universe and quantum gravity. Searches for PBHs serve as a direct probe the spectrum of density perturbations in the early universe on a scale currently inaccessible to any other direct form of observation (\citet{2005astro.ph.11743C}).  This provides crucial input to models of early cosmology, specifically those involving the inflationary paradigm~(\citet{2005astro.ph.11743C}).  Constraints on the topological phase-transition give important input into the allowed compactification scale in extra dimensional models.  Such results, while serving as a good complement to experimental results from accelerator based searches for extra spatial dimensions, also probe a compactification scale that is inaccessible to even the most sensitive accelerator based experiments.  In the context of universal extra dimensions (UED), for example, the bound set here improves upon the current threshold set by the Large Hadron Collider (LHC) by nearly an order of magnitude~(\citet{2012PhLB..718..411A}).
However, it must be noted that this limits presupposes PBHs exist in the appropriate mass range. Thus any limit would apply to extra dimensional models that could be demonstrated to also lead generically to PBH production at the requisite mass scale when applied to early cosmology.

We found an upper limit to the rate density of PBH explosions of $r \approx 4.2 \times 10^{- 7}\,\rm pc^{-3} \,yr^{-1}$ for the standard scenario.
It is useful to compare this result with results from previous surveys.
Typical rate limits for the results analyzed by Phinney and
Taylor, at observing frequencies below $\sim 100$ MHz,
are $r \approx 10^{- 6} \, \rm{pc^{- 3}  \,yr^{- 1}}$ where they assumed $\eta \approx 1$ and that PBHs are uniformly
distributed throughout space~({\citet{Phinney:1979})}. Their analysis of observations from the Arecibo
Observatory produced a rate limit $r \approx 2 \times 10^{- 9} \, \rm{pc^{- 3} \,yr^{- 1}}$. However, they assumed that the
spectrum of a pulse from an exploding PBH is flat over the frequency ranged
covered by their survey. Blandford's analysis~(\citet{Blandford:1977}) shows
that this is not the case, and that the spectrum falls dramatically above the
critical frequency.

Using Blandford's spectrum~(\citet{Blandford:1977}) and the observational parameters from Phinney and Taylor~(\citet{Phinney:1979})
we determine a rate for these Arecibo observations paralleling the calculation detailed in section~\ref{sec:limit} for ETA. The only differences between the ETA analysis and the Arecibo analysis are (1)~$N_{dipole} = 2$ for Arecibo, in
equation~(\ref{eq:SN}), and (2)~a careful analysis of equation~(\ref{eq:SN})
shows that the Arecibo observations are most sensitive to PBH explosions with
$\gamma_f = 10^{4.6}$. The resulting upper limit to the rate of exploding
PBHs is $r \approx 1.1 \times 10^{- 8}\,\rm pc^{-3} \,yr^{-1}$. The rate limit for $\gamma_f=10^{4.3}$ is $4.8 \times 10^{- 6}\,\rm pc^{-3} \,yr^{-1}$ which is more than an order of magnitude less stringent than the rate limit presented for our observations, $r \approx 2.3 \times 10^{-7} \,\rm{pc}^{-3}\,\rm{y}^{-1}$. The Arecibo observations had a duration of $\sim 300$~hours and our ETA observations had a duration of only 4.15~hours. That the search described here can set a stronger constraint for $\gamma_f=10^{4.3}$ than the Arecibo observations is due in part to the fact that the critical frequency related to this fireball Lorentz factor is below the observing frequency of Arecibo. As noted above the intensity for emission from a PBH explosion falls precipitously above the critical frequency. Thus it is expected that instruments that observe at different frequencies should be most effective at probing different sections of the $\gamma_f$ parameter space.  

These points make manifest the relative
merits of the two possible search strategies (large single dish with a small
solid angle versus an array of dipoles with a large solid angle) which is a primary motivation for the observations presented here.
In the case of PBH explosions, the dipole array
strategy is more efficient --- particularly since future observations can
easily expand upon the duration of observations with an array such as LWA1 and LOFAR. Such instruments possess observational capabilities that the ETA lacks, including the ability to phase their dipole arrays to form beams and the ability to conduct all-sky imaging. The ability to form beams could be quite useful in searching for PBH explosions even though such an observing strategy necessarily reduces the solid angle observed. This is because the increase in the distance observed would allow for an effective search of the galactic halo while still retaining a relatively large observing solid angle. 
All-sky imaging could in principle be used to search for PBH explosions but is inherently more problematic because it can be very difficult to obtain the $DM$ for an observed signal using such a method. Without this information, eliminating terrestrial-based signals becomes a greater challenge. Also determining the $DM$ of a signal allows for an estimate of the distance to the source which in turn allows for a determination of the energy released by the source. This information is critical for establishing that the source was a PBH explosion and for distinguishing between the two PBH explosion scenarios described above.


It is also useful to compare these results with (\citet{Benz&Paesold:1998}), who utilize the spectrum from (\citet{Blandford:1977}) and take account of different possible $\gamma_f$ for the expanding shell of charged particles.
They are most sensitive for a PBH explosion with $\gamma_f \approx 10^{4.6}$, which they can detect, with $S/N > 5$ out to a distance of 70~pc.
For the nominal source parameters noted above and an effective observing time of 0.53 years, the upper
limit they set is $r = 4.8 \times 10^{- 3} \,\rm{pc^{- 3} \,yr^{- 1}}$ for $\eta \approx 1$ and $\gamma_f \approx
10^{4.6}$, assuming that PBHs are uniformly distributed throughout space. The observations presented here set a more stringent constraint for all possible values of $\gamma_f$
with only $\approx 4$ hours of observing.

It has been suggested that PBHs may contribute to the dark matter in our
Galaxy~{(\citet{1991ApJ...371..447M, ivanov1994inflation, afshordi2003primordial, seto2004search, abramowicz2009no})}.
PBHs would, therefore, not be uniformly distributed throughout space, but
clumped together in the halos of galaxies~{(\citet{1996ApJ...459..487W,abramowicz2009no,lehoucq2009new})}.
(\citet{lehoucq2009new}) used whole-sky survey data obtained by the Energetic
Gamma-Ray Experiment Telescope (EGRET) to calculate the PBH explosion rate.
They assumed PBHs to be distributed as dark matter and to have an initial mass
$M = 5 \times 10^{14}$~g.
By comparing the predicted cumulative Galactic $\gamma$-ray emission to the one observed by the EGRET satellite, they found
an upper limit to the local rate of PBH explosions ($\eta = 1$) to be $\approx
0.059 \, \rm{pc^{- 3}\, yr^{- 1}}$.
The limit calculated in the current work, using similar assumptions, is about 5
orders of magnitude lower.


\section{Acknowledgements}
We thanks Andy O'Bannon for a thorough and insightful reading of this manuscript and Jamie Tsai for thoughtful commentary. Part of this research was carried out while the primary author held a NRC Postdoctoral Research Appointment at the U.S. Naval Research Laboratory.  M. Kavic and A. Larracuente were supported by NASA Grant NNX11AI27G.  Construction and operation of ETA was supported through an NSF Advanced Techniques and Instrumentation (ATI) grant No.~AST-0504677, and was hosted at the Pisgah Astronomical Research Institute (PARI) near Balsam Grove, NC from 2005 through 2008.  Basic research in radio astronomy at NRL is supported by 6.1 base funding.



\begin{thebibliography}{}
\expandafter\ifx\csname natexlab\endcsname\relax\def\natexlab#1{#1}\fi

\bibitem[{Abramowicz {et~al.}(2009)Abramowicz, Becker, Biermann, Garzilli,
  Johansson, \& Qian}]{abramowicz2009no}
Abramowicz, M.~A., Becker, J.~K., Biermann, P.~L., {et~al.} 2009, The
  Astrophysical Journal, 705, 659

\bibitem[{Afshordi {et~al.}(2003)Afshordi, McDonald, \&
  Spergel}]{afshordi2003primordial}
Afshordi, N., McDonald, P., \& Spergel, D. 2003, The Astrophysical Journal
  Letters, 594, L71

\bibitem[{{ATLAS Collaboration Aad} {et~al.}(2012){ATLAS Collaboration Aad},
  {Abajyan}, {Abbott}, {Abdallah}, {Abdel Khalek}, {Abdelalim}, {Abdinov},
  {Aben}, {Abi}, {Abolins}, \& et~al.}]{2012PhLB..718..411A}
{ATLAS Collaboration Aad}, G., {Abajyan}, T., {Abbott}, B., {et~al.} 2012,
  Physics Letters B, 718, 411

\bibitem[{Balsano {et~al.}(1996)Balsano, Thorsett, Coles, Ray, Rhodes, Rickett,
  Barthelmy, Butterworth, Cline, Gehrels, {et~al.}}]{balsano1996searching}
Balsano, R., Thorsett, S., Coles, W., {et~al.} 1996, in AIP Conference
  Proceedings, Citeseer, 575--579

\bibitem[{Barrau {et~al.}(2014)Barrau, Rovelli, \& Vidotto}]{barrau2014fast}
Barrau, A., Rovelli, C., \& Vidotto, F. 2014, Physical Review D, 90, 127503


\bibitem[{Belotsky {et~al.}(2014)Belotsky, Dmitriev, Esipova, Gani, Grobov,
  Khlopov, Kirillov, Rubin, \& Svadkovsky}]{Belotsky:2014kca}
Belotsky, K.~M., Dmitriev, A.~D., Esipova, E.~A., {et~al.} 2014, Mod. Phys.
  Lett., A29, 1440005


\bibitem[{{Benz} \& {Paesold}(1998)}]{Benz&Paesold:1998}
{Benz}, A.~O., \& {Paesold}, G. 1998, Astron. and Astrophys., 329, 61

\bibitem[{{Blandford}(1977)}]{Blandford:1977}
{Blandford}, R.~D. 1977, Mon. Not. R. Astron. Soc., 181, 489

\bibitem[{Broniowski {et~al.}(2004)Broniowski, Florkowski, \&
  Glozman}]{Broniowski:2004yh}
Broniowski, W., Florkowski, W., \& Glozman, L.~Y. 2004, Phys.Rev., D70, 117503

\bibitem[{{Cane}(1979)}]{1979MNRAS.189..465C}
{Cane}, H.~V. 1979, Monthly Notices of the Royal Astronomical Society, 189, 465

\bibitem[{{Carr}(2005)}]{2005astro.ph.11743C}
{Carr}, B.~J. 2005, ArXiv Astrophysics e-prints, arXiv:astro-ph/0511743

\bibitem[{Cohen \& Krejcirik(2012)}]{Cohen:2011cr}
Cohen, T.~D., \& Krejcirik, V. 2012, J.Phys., G39, 055001

\bibitem[{{Cordes} \& {McLaughlin}(2003)}]{Cordes&McLaughlin:2003}
{Cordes}, J.~M., \& {McLaughlin}, M.~A. 2003, Astrophys. J., 596, 1142

\bibitem[{{Cordes} \& {Rickett}(1998)}]{1998ApJ...507..846C}
{Cordes}, J.~M., \& {Rickett}, B.~J. 1998, \apj, 507, 846

\bibitem[{Cutchin(2011)}]{cutchin2011search}
Cutchin, S.~E. 2011, PhD thesis, Virginia Polytechnic Institute and State
  University

\bibitem[{Deshpande(2009)}]{KshitijaETA:2009ps}
Deshpande, K. 2009, {A Dedicated Search for Low Frequency Radio Transient
  Astrophysical Events using ETA}, m.S. Thesis, Virginia Polytechnic Inst \&
  State Univ.

\bibitem[{{Ellingson} {et~al.}(2007){Ellingson}, {Simonetti}, \&
  {Patterson}}]{Ellingson:2007}
{Ellingson}, S.~W., {Simonetti}, J.~H., \& {Patterson}, C.~D. 2007, IEEE Trans.
  Antenn. and Propag., 55, 826

\bibitem[{Gregory \& Laflamme(1993)}]{gregory1993black}
Gregory, R., \& Laflamme, R. 1993, Physical review letters, 70, 2837

\bibitem[{Haslam {et~al.}(1982)Haslam, Salter, Stoffel, \&
  Wilson}]{Haslam:1982}
Haslam, C., Salter, C., Stoffel, H., \& Wilson, W. 1982,
  Astron.Astrophys.Suppl.Ser., 47, 1

\bibitem[{{Hawking}(1975)}]{Hawking:1975}
{Hawking}, S.~W. 1975, Communications in Mathematical Physics, 43, 199

\bibitem[{Ivanov {et~al.}(1994)Ivanov, Naselsky, \&
  Novikov}]{ivanov1994inflation}
Ivanov, P., Naselsky, P., \& Novikov, I. 1994, Physical Review D, 50, 7173

\bibitem[{Jelley {et~al.}(1965)Jelley, Fruin, Porter, Weekes, Smith, \&
  Porter}]{jelley1965radio}
Jelley, J., Fruin, J., Porter, N., {et~al.} 1965

\bibitem[{Kavic {et~al.}(2008)Kavic, Simonetti, Cutchin, Ellingson, \&
  Patterson}]{Kavic:2008}
Kavic, M., Simonetti, J.~H., Cutchin, S.~E., Ellingson, S.~W., \& Patterson,
  C.~D. 2008, JCAP, 0811, 017




\bibitem[{Khlopov {et~al.}(1985)Khlopov, Malomed, \&
  Zeldovich}]{Khlopov:1985jw}
Khlopov, M., Malomed, B.~A., \& Zeldovich, I.~B. 1985, Mon. Not. Roy. Astron.
  Soc., 215, 575

\bibitem[{Khlopov(2010)}]{Khlopov:2008qy}
Khlopov, M.~{\relax Yu}. 2010, Res. Astron. Astrophys., 10, 495


\bibitem[{{Kol}(2002)}]{Kol:2002}
{Kol}, B. 2002, arXiv:hep-ph/0207037, ArXiv High Energy Physics - Phenomenology
  e-prints

\bibitem[{Lehoucq {et~al.}(2009)Lehoucq, Cass{\'e}, Casandjian, \&
  Grenier}]{lehoucq2009new}
Lehoucq, R., Cass{\'e}, M., Casandjian, J.-M., \& Grenier, I. 2009, arXiv
  preprint arXiv:0906.1648

\bibitem[{{MacGibbon} \& {Carr}(1991)}]{1991ApJ...371..447M}
{MacGibbon}, J.~H., \& {Carr}, B.~J. 1991, ApJ, 371, 447

\bibitem[{McLaughlin {et~al.}(2006)}]{McLaughlin:2006}
McLaughlin, M.~A., {et~al.} 2006, Nature, 439, 817

\bibitem[{{Meikle} \& {Colgate}(1978)}]{Meikle&Colgate:1978}
{Meikle}, W.~P.~S., \& {Colgate}, S.~A. 1978, Astrophys. J., 220, 1076

\bibitem[{{Page} \& {Hawking}(1976)}]{Page:1976}
{Page}, D.~N., \& {Hawking}, S.~W. 1976, Astrophys. J., 206, 1

\bibitem[{{Phinney} \& {Taylor}(1979)}]{Phinney:1979}
{Phinney}, S., \& {Taylor}, J.~H. 1979, Nature, 277, 117

\bibitem[{Porter {et~al.}(1965)Porter, Long, McBreen, Murnaghan, \&
  Weekes}]{porter1965radio}
Porter, N., Long, C., McBreen, B., Murnaghan, D., \& Weekes, T. 1965, in
  International Cosmic Ray Conference, Vol.~2, 706

\bibitem[{{Rees}(1977)}]{Rees:1977}
{Rees}, M.~J. 1977, Nature, 266, 333

\bibitem[{Seto \& Cooray(2004)}]{seto2004search}
Seto, N., \& Cooray, A. 2004, Physical Review D, 70, 063512

\bibitem[{Stappers {et~al.}(2011)Stappers, Hessels, Alexov, Anderson, Coenen,
  Hassall, Karastergiou, Kondratiev, Kramer, van Leeuwen,
  {et~al.}}]{stappers2011observing}
Stappers, B., Hessels, J., Alexov, A., {et~al.} 2011, Astronomy \&
  Astrophysics, 530, A80

\bibitem[{{Taylor} {et~al.}(2012){Taylor}, {Ellingson}, {Kassim}, {Craig},
  {Dowell}, {Wolfe}, {Hartman}, {Bernardi}, {Clarke}, {Cohen}, {Dalal},
  {Erickson}, {Hicks}, {Greenhill}, {Jacoby}, {Lane}, {Lazio}, {Mitchell},
  {Navarro}, {Ord}, {Pihlstr{\"o}m}, {Polisensky}, {Ray}, {Rickard},
  {Schinzel}, {Schmitt}, {Sigman}, {Soriano}, {Stewart}, {Stovall}, {Tremblay},
  {Wang}, {Weiler}, {White}, \& {Wood}}]{2012JAI.....150004T}
{Taylor}, G.~B., {Ellingson}, S.~W., {Kassim}, N.~E., {et~al.} 2012, Journal of
  Astronomical Instrumentation, 1, 50004

\bibitem[{{Thornton} {et~al.}(2013){Thornton}, {Stappers}, {Bailes},
  {Barsdell}, {Bates}, {Bhat}, {Burgay}, {Burke-Spolaor}, {Champion}, {Coster},
  {D'Amico}, {Jameson}, {Johnston}, {Keith}, {Kramer}, {Levin}, {Milia}, {Ng},
  {Possenti}, \& {van Straten}}]{2013Sci...341...53T}
{Thornton}, D., {Stappers}, B., {Bailes}, M., {et~al.} 2013, Science, 341, 53

\bibitem[{Tingay {et~al.}(2013)Tingay, Goeke, Bowman, Emrich, Ord, Mitchell,
  Morales, Booler, Crosse, Wayth, {et~al.}}]{tingay2013murchison}
Tingay, S., Goeke, R., Bowman, J.~D., {et~al.} 2013, Publications of the
  Astronomical Society of Australia, 30, e007

\bibitem[{{Wright}(1996)}]{1996ApJ...459..487W}
{Wright}, E.~L. 1996, The Astrophysical Journal, 459, 487

\end{thebibliography}

\end{document}